\documentclass[a4paper,11pt]{article}
\usepackage{jheppub} % for details on the use of the package, please
\usepackage{multirow}

\newcommand{\GeV}{\mbox{GeV}} \newcommand{\MeV}{\mbox{MeV}}
\newcommand{\Br}{\mathrm{Br}}
 
\newcommand{\R}{\mathcal{R}}
\newcommand{\M}{\mathcal{M}}
\newcommand{\HH}{\mathcal{H}}

\title{\boldmath Production of $K$ mesons in exclusive $B_{c}$ decays}
\author{A.V. Luchinsky}
\affiliation{Institute for High Energy Physics, Protvino, Russia}
\emailAdd{Alexey.Luchinsky@ihep.ru}

\abstract{
The paper is devoted to investigation of $K$ mesons production in
exclusive $B_{c}$ decays $B_{c}\to\psi^{(')}K^{+}K^{-}\pi^{+}$ and
$B_{c}\to\psi^{(')}K^{+}\pi^{+}\pi^{-}$. In the framework of resonance
approximation we obtain  theoretical predictions for branching
fractions of these decays and present distributions over different
kinematical variables.
}

\begin{document}

\maketitle
\flushbottom

\section{Introduction}

Recently there was a large progress in theoretical and experimental
study of $B_{c}$ mesons. In addition to measurement of its mass and
lifetime, which is in excellent agreement with theoretical predictions,
the branching fractions of some decays were also determined. One can
mention, for example, semileptonic decays $B_{c}\to J/\psi\mu\nu$
and exclusive decays into vector charmonium and a system of light
mesons, i.e. $B_{c}\to J/\psi\pi$, $B_{c}\to\psi(2S)\pi$, $B_{c}\to J/\psi+3\pi$,
etc. Production of $K$-mesons in exclusive $B_{c}$ decays, however,
was not yet considered. Below we give theoretical analysis of
these reactions.

\section{Exclusive $B_c$ decays}

\begin{figure}
\begin{centering}
\includegraphics[width=0.7\textwidth]{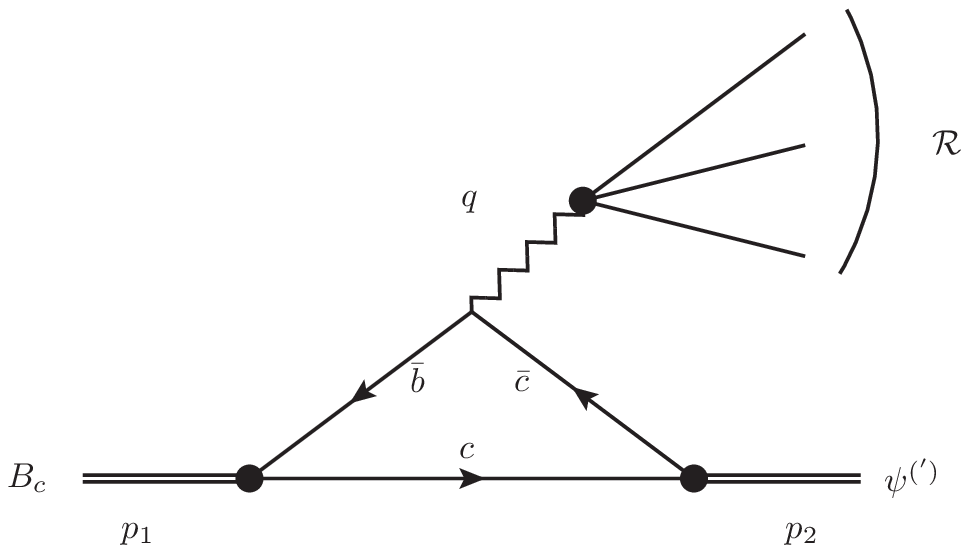}
\par\end{centering}

\caption{Feynman diagram for $B_{c}\to\psi^{(')}+\R$ decay\label{diag:bc}}
\end{figure}

Exclusive $B_{c}$ meson decays into charmonia and semileptonic pair
or a set of light mesons take special place. On the partonic level
in the factorization approximation such processes proceed via weak
decay of $b$ quark into $c$. The virtual $W$ boson afterwards hadronizes
into final system of light particles, while $c\bar{c}$ pair transforms
into charmonium meson. Typical diagram of such process is shown in
Fig.\ref{diag:bc}. The amplitude of the reaction can be written
in the form
\begin{eqnarray}
\M\left(B_{c}\to\psi^{(')}+\R\right) & = & \frac{G_{F}V_{bc}}{\sqrt{2}}a_{1}(m_{b})\HH^{\mu}\epsilon_{\mu}^{(\R)},\label{eq:amp}
\end{eqnarray}
where $G_{F}$ is the Fermi constant, $V_{bc}$ is the quark mixing
matrix element, Wilson coefficient $a_{1}(m_{b})\approx1.14$ describes 
the effect of QCD corrections \cite{Buchalla:1995vs}, while
vectors $\HH^{\mu}$ and $\epsilon_{\mu}^{(\R)}$ describe $B_{c}\to\psi^{(')}W$
and $W\to\R$ transitions.

Let us first discuss the first vertex. It is usually written in terms
of $B_{c}$ formfactors and one of popular parameterizations has the
form
\begin{eqnarray}
\HH^{\mu} & = & 2M_{2}A_{0}\left(q^{2}\right)\frac{q^{\mu}\left(\epsilon q\right)}{q^{2}}+
\left(M_{1}+M_{2}\right)A_{1}\left(q^{2}\right)
	\left(\epsilon^{\mu}-\frac{q^{\mu}\left(q\epsilon\right)}{q^{2}}\right)-
\nonumber \\ &  & 
A_{2}\left(q^{2}\right)\frac{\left(q\epsilon\right)}{M_{1}+M_{2}}\left(p_{1}+p_{2}-\frac{M_{1}^{2}-M_{2}^{2}}{q^{2}}\right)-
\nonumber \\ &  & 
\frac{2iV\left(q^{2}\right)}{M_{1}+M_{2}}e_{\mu\nu\alpha\beta}\epsilon^{\nu}p_{1}^{\alpha}p_{2}^{\beta},
\end{eqnarray}
where $p_{1,2}$ and $M_{1,2}$ are momenta and masses of initial
and final heavy quarkonia, $q=p_{1}-p_{2}$ is the transferred momentum,
$\epsilon_\mu$ is the polarization vector of $\psi^{(')}$ meson, and
$A_{0,1,2}\left(q^{2}\right),$ $V(q^{2})$ are the formfactors. It
is clear that these functions cannot be determined from perturbative
theory, so some other approach should be used for example
QCD sum rules \cite{Huang:2007kb,Kiselev:1999sc,Kiselev:2000nf,Kiselev:2000pp,Kiselev:2002vz},
different potential quark models \cite{Kiselev:1992tx,Gershtein:1994jw,Gershtein:1997qy,Colangelo:1999zn,Ivanov:2005fd},
light-front models \cite{Anisimov:1998xv,Choi:2009ym,Choi:2009ai},
etc. In our paper we will use formfactors stets presented in papers
\cite{Kiselev:2000pp} and \cite{Ebert:2003cn} (in the following
they will be labeled as SR and PM respectively). In table \ref{tab:FF}
we show values of these formfactors at different points.

\begin{table}
\begin{centering}
\begin{tabular}{|c|c|c|c|c||c|c|c|c|}
\hline 
 & \multicolumn{4}{c||}{$B_{c}\to J/\psi$} & \multicolumn{4}{c|}{$B_{c}\to\psi(2S)$}\tabularnewline
\hline 
 & \multicolumn{2}{c|}{SR} & \multicolumn{2}{c||}{PM} & \multicolumn{2}{c|}{SR} & \multicolumn{2}{c|}{PM}\tabularnewline
\hline 
 & $q^{2}=0$  & $q^{2}=q_{\max}^{2}$ & $q^{2}=0$ & $q^{2}=q_{\max}^{2}$ & $q^{2}=0$  & $q^{2}=q_{\max}^{2}$ & $q^{2}=0$ & $q^{2}=q_{\max}^{2}$\tabularnewline
\hline 
\hline 
$A_{0}$ & $0.60$ & $1.6$ & $0.42$ & $1.0$ & $0.15$ & $0.28$ & $0.24$ & $0.011$\tabularnewline
\hline 
$A_{1}$ & $0.63$ & $1.3$ & $0.5$ & $0.87$ & $0.14$ & $0.21$ & $0.17$ & $-0.0069$\tabularnewline
\hline 
$A_{2}$ & $0.69$ & $1.4$ & $0.73$ & $1.3$ & $0.13$ & $0.19$ & $0.14$ & $0.65$\tabularnewline
\hline 
$V$ & $1.0$ & $2.1$ & $0.49$ & $1.3$ & $0.3$ & $0.44$ & $0.24$ & $-0.33$\tabularnewline
\hline 
\end{tabular}
\par\end{centering}

\caption{$B_{c}\to J/\psi W$ formfactors \label{tab:FF}}
\end{table}

From the presented in this table data  it is
clear that in the case of $B_{c}\to J/\psi$ transition $q^{2}$ dependence
and of formfactors are similar for different models, the main difference
is in the overall normalization. In the case of $B_{c}\to\psi(2S)$
transition, on the contrary, the form of distribution depends strongly
on the choice of the model. This difference is caused by the fact,
that the wave function of excited charmonium meson should have a node
that is not represented in QCD sum rules results. The role of this
node is discussed, for example, in paper \cite{Luchinsky:2012rk}.

The last factor in eq. (\ref{eq:amp}), i.e. the amplitude of $W\to\R$
transition, is also nonperturbative. In our paper we shall use resonance
approximation that gives good results in the case of $B_{c}\to J/\psi+3\pi$
decay \cite{Likhoded:2009ib,LHCb:2012ag}. The same method was used
later in papers \cite{Likhoded:2010jr,Berezhnoy:2011is,Luchinsky:2012rk}
to study some other exclusive decays of $B_{c}$ meson. In this approach
the amplitude of the process is written as an amplitude of subsequent
decays of virtual resonances with suitable quantum numbers. For example,
in the case of $B_{c}\to J/\psi+3\pi$ decay the reaction $W\to a_{1}\to\rho\pi\to3\pi$
was used. For reactions considered in the present paper we take into
account also contributions of $K_{1}(1270)$, $K_{1}(1400)$, and
$K^{*}$ mesons.

\begin{table}
\begin{centering}
\begin{tabular}{|c|c|c|c|c|}
\hline 
\multirow{2}{*}{$\R$} & \multicolumn{2}{c|}{$B_{c}\to J/\psi+\R$} & \multicolumn{2}{c|}{$B_{c}\to\psi(2S)+\R$}\tabularnewline
\cline{2-5} 
 & SM & PM & SM & PM\tabularnewline
\hline 
\hline 
$\pi$ & $0.17$ & $0.064$ & $0.0066$ & $0.014$\tabularnewline
\hline 
$K\pi\pi$ & $0.03$ & $0.011$ & $0.0011$ & $0.00064$\tabularnewline
\hline 
$KK\pi$ & $0.081$ & $0.03$ & $0.0023$ & $0.00073$\tabularnewline
\hline 
\end{tabular}
\par\end{centering}

\caption{Branching fractions of exclusive $B_{c}$ decays (in \%)\label{tab:Br}}
\end{table}

%Experimental results are often reported as branching fractions normalized to branching fraction of $B_{c}\to\psi^{(')}\pi$ decay.
The branching fractions of $B_c$ decays are often normalized to branching fraction of $B_c\to\psi^{(')}\pi$ reactions.
The latter
quantity can easily calculated using presented above expressions with
effective polarization vector $\epsilon_{\mu}^{(\pi)}$ equal to
\begin{eqnarray}
\epsilon_{\mu}^{(\pi)} & = & f_{\pi}k_{\mu}\delta\left(q^{2}-m_{\pi}^{2}\right).
\end{eqnarray}
Here $k_{\mu}$ and $m_{\pi}=140\,\MeV$ are momentum and mass of
produced pion, while $f_{\pi}\approx130\,\MeV$ is its mesonic constant.
Numerical values of branching fractions of $B_{c}\to\psi^{(')}\pi$
decays for different sets of formfactors are shown in table \ref{tab:Br}.

\section{$B_c\to \psi^{(')}+K\pi\pi$}

\begin{figure}
\begin{centering}
\includegraphics{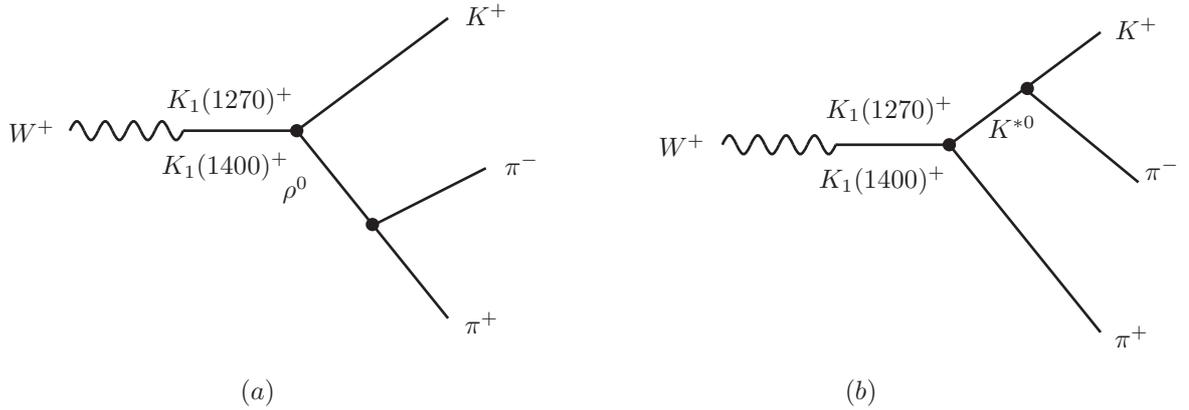}
\par\end{centering}

\caption{Typical diagrams for $W\to K\pi\pi$ transition\label{diag:Kpp}}
\end{figure}

\begin{figure}
\begin{centering}
\includegraphics[width=\textwidth]{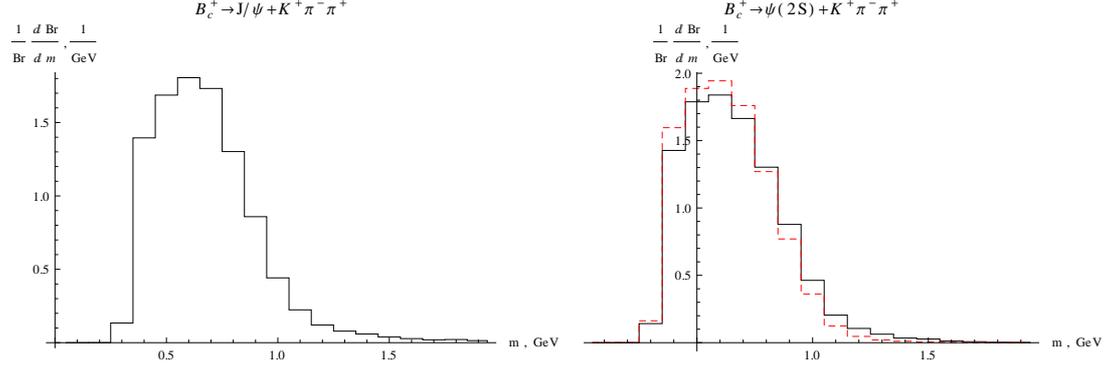}
\par\end{centering}

\caption{Distribution over $m_{\pi^{+}\pi^{-}}$ in the case of $B_{c}\to J/\psi+K\pi\pi$
(left figure) and $B_{c}\to\psi(2S)+K\pi\pi$ (right figure) decays.
Solid and dashed lines  correspond to SR and PM formfactor
sets respectively\label{fig:Kpp12}}

\end{figure}

\begin{figure}
\begin{centering}
\includegraphics[width=\textwidth]{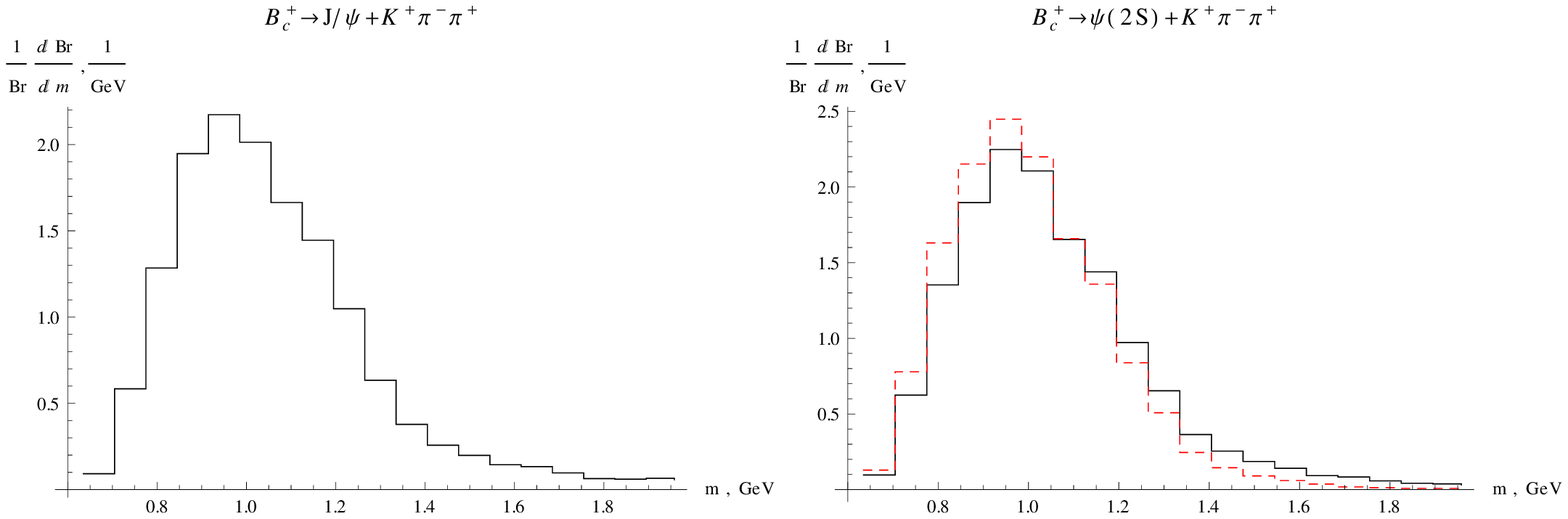}
\par\end{centering}

\caption{Distribution over $m_{K^{+}\pi^{+}}$ in the case of $B_{c}\to\psi^{(')}+K\pi\pi$
decay. Notations are same as in Fig.\ref{fig:Kpp12}\label{fig:KppK1}}
\end{figure}

\begin{figure}
\begin{centering}
\includegraphics[width=\textwidth]{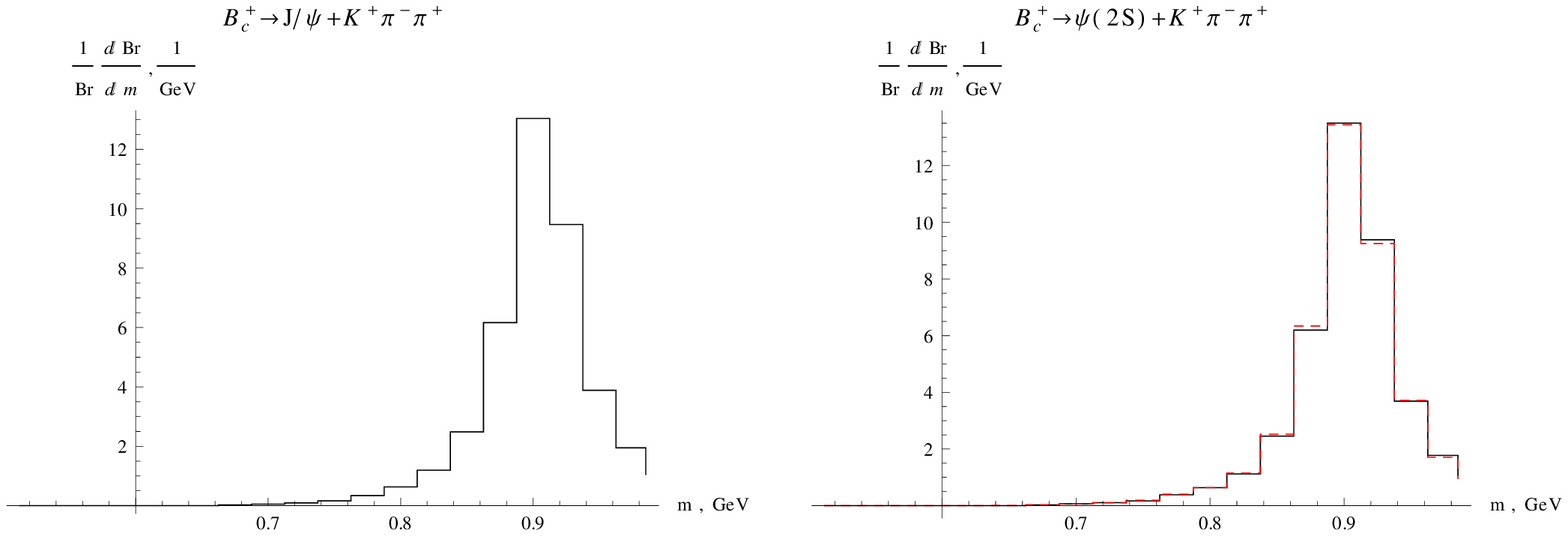}
\par\end{centering}

\caption{Distribution over $m_{K^{+}\pi^{-}}$ in the case of $B_{c}\to\psi^{(')}+K\pi\pi$
decay. Notations are same as in Fig.\ref{fig:Kpp12}\label{fig:KppK2}}
\end{figure}

Let us first consider the $B_{c}\to\psi^{(')}+K\pi\pi$ decay. In
this case typical diagrams for $W\to K\pi\pi$ transition are shown
in Fig.\ref{diag:Kpp}. From quantum numbers of virtual resonances
one can determine the amplitude of this reaction:
\begin{eqnarray}
\epsilon_{\mu}^{(K\pi\pi)} & = & g_{K(1270)\rho}D_{K(1270)}(q)D_{\rho}(k_{1}+k_{2})(k_{1}-k_{2})_{\mu}+
\nonumber \\ &&
g_{K(1270)K^{*}}D_{K(1270)}(q)D_{K^{*}}(k_{2}+k_{3})(k_{2}-k_{3})_{\mu}+
\nonumber \\ &&
  (K_{1}(1270)\to K_{1}(1400)),
\end{eqnarray}
where $k_{1,2,3}$ are momenta of $\pi^{+}$, $\pi^{-}$ and $K^{+}$
mesons respectively and we introduce the shorthand notation for meson
propagator
\begin{eqnarray}
D_{A}(p) & = & \frac{1}{p^{2}-M_{A}^{2}+iM_{A}\Gamma_{A}}.
\end{eqnarray}
Coupling constants $g_{K\rho},$ $g_{KK^{*}}$ can be
determined from analysis of corresponding decays of $K$ mesons, but
we prefer to use directly experimental $q^2$ distributions from $\tau\to\nu_{\tau}+KK\pi$
decay, as it was done for $B_{c}\to J/\psi+3\pi$ reaction in papers
\cite{Likhoded:2009ib,Kuhn:1990ad}. From the fit of presented in
ref.\cite{Lee:2010tc} data we obtain the following values of these
constants:
\begin{eqnarray}
g_{K(1270)\rho} & = & -4.14\times10^{-2}\,\GeV^{3},\qquad g_{K(1270)K^{*}}=0.17\,\GeV^{3},\\
g_{K(1400)\rho} & = & 0.13\,\GeV^{3},\qquad g_{K(1400)K^{*}}=0.24\,\GeV^{3}.
\end{eqnarray}
%In Fig. we show the $q^{2}$ distributions for $\tau\to\nu_{\tau}+K\pi\pi$
%decay in comparison with results from paper \cite{Lee:2010tc}.
 Using
presented above expressions one can calculate the branching fractions
of $B_{c}\to\psi^{(')}K^{+}\pi^{+}\pi^{-}$ decays. Our results for
different choice of formfactor set are shown in the third row of table
\ref{tab:Br}. After normalization to $B_{c}\to\psi^{(')}\pi$ branching
fractions we obtain the ratios
\begin{eqnarray}
\frac{\Br\left(B_{c}\to J/\psi+K\pi\pi\right)}{\Br\left(B_{c}\to J/\psi\pi\right)} & = & 0.18, 
\frac{\Br\left(B_{c}\to\psi(2S)+K\pi\pi\right)}{\Br\left(B_{c}\to\psi(2S)\pi\right)}  =  0.16
\end{eqnarray}
for SR model of form factors and
\begin{eqnarray}
\frac{\Br\left(B_{c}\to J/\psi+K\pi\pi\right)}{\Br\left(B_{c}\to J/\psi\pi\right)} = 0.17,
\frac{\Br\left(B_{c}\to\psi(2S)+K\pi\pi\right)}{\Br\left(B_{c}\to\psi(2S)\pi\right)} = 0.047
\end{eqnarray}
for PM.
Distributions over invariant masses of $K\pi$
and $\pi\pi$ pairs are presented in Fig.\ref{fig:Kpp12}, \ref{fig:KppK1},
\ref{fig:KppK2}. A noticable peak in the last figure is caused by $K^*$ meson contribution (see diagram \ref{diag:Kpp}b). 
One can easily see that in the case of $J/\psi$
meson in the final state branching fractions ratio and forms of distributions
do not depend on the choice of formfactors model, while in the case
of $B_{c}\to\psi(2S)+K\pi\pi$ decays modification of formfactor's
set changes theoretical predictions significantly. 

\section{$B_c\to\psi^{(')}+KK\pi$}

\begin{figure}
\begin{centering}
\includegraphics[width=\textwidth]{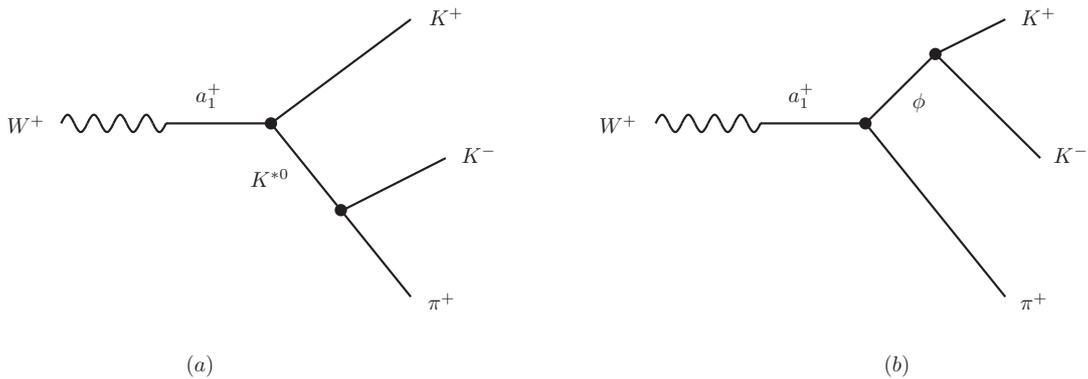}
\par\end{centering}

\caption{Typical diagrams for $W\to KK\pi$ transition\label{diag:KKpi}}

\end{figure}

\begin{figure}
\begin{centering}
\includegraphics[width=\textwidth]{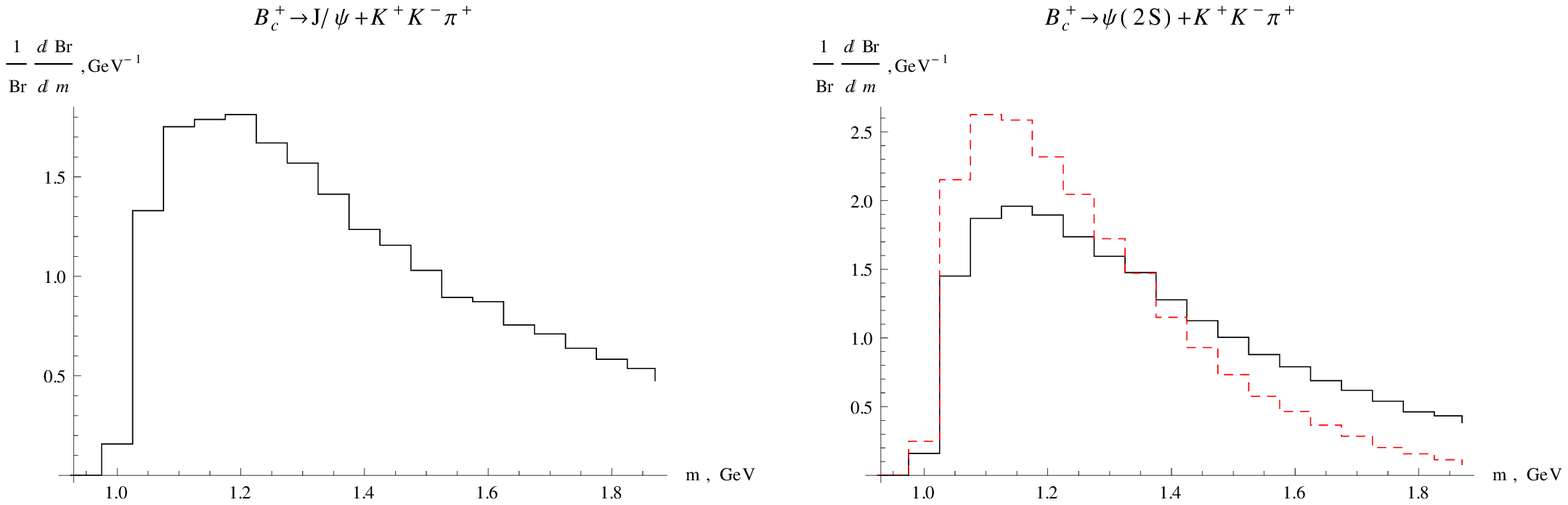}
\par\end{centering}

\caption{Distribution over $m_{K^{+}K^{-}}$ in the case of $B_{c}\to\psi^{(')}+KK\pi$
decays. Notations are same as in Fig.\ref{fig:Kpp12}\label{fig:KKp12}}
\end{figure}

\begin{figure}
\begin{centering}
\includegraphics[width=\textwidth]{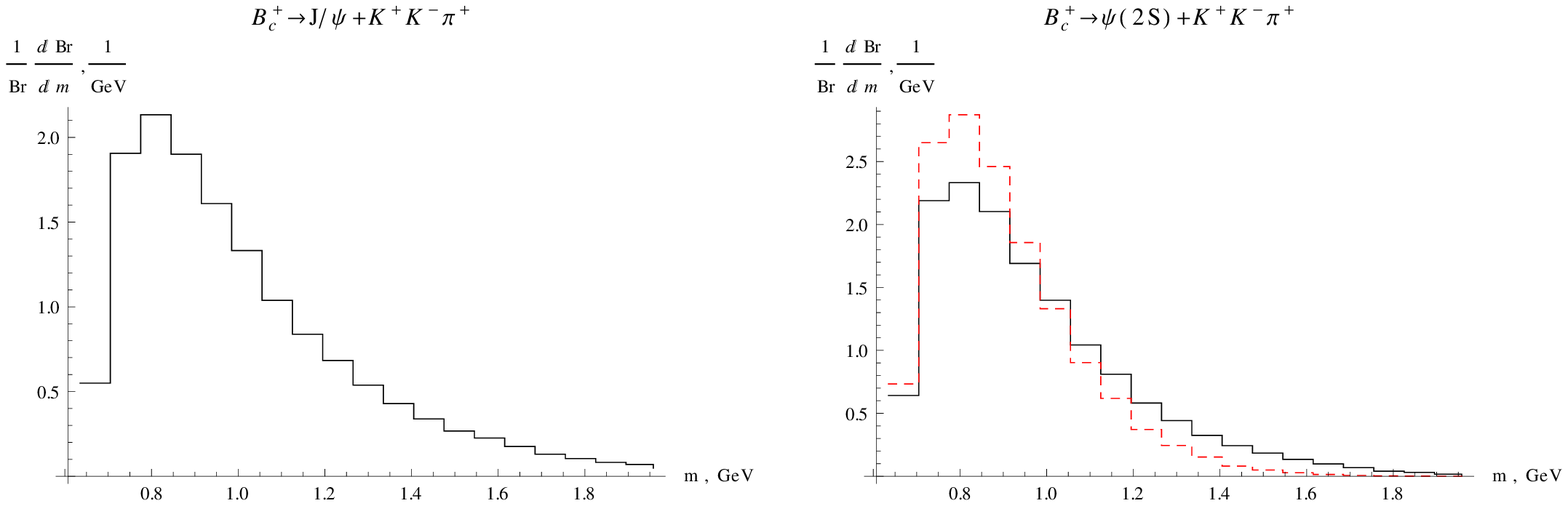}
\par\end{centering}

\caption{Distribution over $m_{K^{+}\pi^{+}}$ in the case of $B_{c}\to\psi^{(')}+KK\pi$
decays. Notations are same as in Fig.\ref{fig:Kpp12}\label{fig:KKp1}}
\end{figure}

\begin{figure}
\begin{centering}
\includegraphics[width=\textwidth]{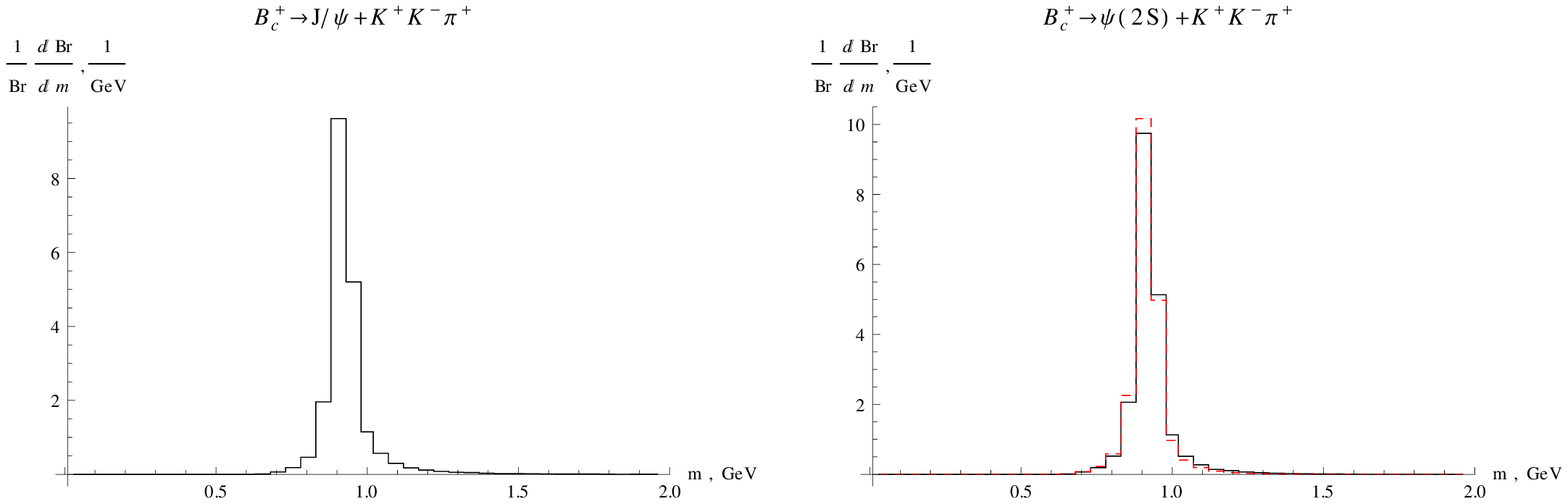}
\par\end{centering}

\caption{Distribution over $m_{K^{+}\pi^{-}}$ in the case of $B_{c}\to\psi^{(')}+KK\pi$
decays. Notations are same as in Fig.\ref{fig:Kpp12}\label{fig:KKp2}}
\end{figure}

Using the same approach one can consider $B_{c}\to\psi^{(')}K^{+}K^{-}\pi^{+}$
decays. In this case decay chains $W\to a_{1}\to KK^{*}\to KK\pi$
and $W\to a_{1}\to\phi\pi\to KK\pi$ can give contributions (see diagrams
shown in Fig.\ref{diag:KKpi}), but the last reaction should be suppressed
due to OZI rule and G-parity violation, so we will not take it into
account. As for diagram shown in Fig.\ref{diag:KKpi}(a), the corresponding
amplitude can be written as
\begin{eqnarray*}
\epsilon_{\mu}^{(KK\pi)} & = & g_{KK\pi}D_{a_{1}}(q)D_{K^{*}}(k_{1}+k_{3})\left(k_{1}-k_{3}\right)_{\mu},
\end{eqnarray*}
where the constant $g_{KK\pi}$
%$g_{KK\pi}\approx\QQQ$
 was determined from $\tau\to\nu_{\tau}+KK\pi$
decay. The branching fractions of $B_{c}\to\psi^{(')}+KK\pi$ decays,
obtained using presented above expressions are listed in the last
row of table \ref{tab:Br}, and the ratios to $\Br\left(B_{c}\to\psi^{(')}\pi\right)$
for different form factor sets are equal to
\begin{eqnarray}
\frac{\Br\left(B_{c}\to J/\psi+KK\pi\right)}{\Br\left(B_{c}\to J/\psi\pi\right)} & = & 0.49,
\frac{\Br\left(B_{c}\to\psi(2S)+KK\pi\right)}{\Br\left(B_{c}\to\psi(2S)\pi\right)}  =  0.34
\end{eqnarray}
for SR set of form factors and
\begin{eqnarray}
\frac{\Br\left(B_{c}\to J/\psi+KK\pi\right)}{\Br\left(B_{c}\to J/\psi\pi\right)} &=& 0.47,
\frac{\Br\left(B_{c}\to\psi(2S)+KK\pi\right)}{\Br\left(B_{c}\to\psi(2S)\pi\right)}  = 0.053
\end{eqnarray}
for PM model. 
It is seen that these values ​​are greater than the corresponding ratios in the case of $B_c\to\psi^{(')}+KK\pi$ decays.
The reason is that the latter processes are suppressed by Cabibbo factor $V_{us}$. 
In our model, this factor is contained in the $W\to K\pi\pi$ transition amplitude. Distributions over invariant masses $m_{K^{+}K^{-}}$, $m_{K^{+}\pi^{+}}$
and $m_{K^{-}\pi^{+}}$ are shown in Fig.\ref{fig:KKp12}, \ref{fig:KKp1},
\ref{fig:KKp2} respectively.


\section{Conclusion}
Let us summarize briefly the results of our note. In the framework of factorization approximation the branching fractions of $B_c\to\psi^{(')}+K\pi\pi$ and $B_c\to\psi^{(')}+KK\pi$ decays were calculated. These branching fractions depend on form factors of $B_c\to\psi^{(')}W$ verticies (in our analysis we use FF models presented in papers \cite{Kiselev:2000pp,Ebert:2003cn}) and amplitudes of $W\to K\pi\pi$, $W\to KK\pi$ transitions. The latter amplitudes were written in the resonance approximation with contributions of $K_1(1270)$,  $K_1(1400)$, $K^*(890)$, $a_1$ and $\rho$ mesons taken into account. Corresponding coupling constants were determined from analysis of $\tau$ lepton decays. We believe that experimental study of discussed in our paper decays can give additional information about both $B_c\to\psi^{(')}$ form factors and physics of light mesons.

\acknowledgments

The author would like to thank A. Likhoded, I. Belyaev and V. Egorychev for useful and fruitful discussions. The work was financially supported by Russian Foundation for Basic Research (grant \#10-00061a), the grant of the President of
Russian Federation (grant \#MK-3513.2012.2), and FRRC grant.

\bibliographystyle{JHEP}
\bibliography{litr}

\end{document}